\documentclass[journal]{IEEEtran}
\ifCLASSINFOpdf
\else
\fi
\usepackage{graphicx}
\usepackage{subfigure}
\usepackage{hyperref}
\usepackage{graphicx}
\usepackage{url}
\usepackage{pslatex}
\usepackage{bm}
\usepackage[comma,numbers,square,sort&compress]{natbib}
\usepackage{amsmath,amssymb,amsthm}  
\usepackage{paralist}

\usepackage{algorithm}
\usepackage{algorithmic}        
\usepackage{multirow}
\renewcommand{\algorithmicrequire}{\textbf{Initialization:}}   
\renewcommand{\algorithmicensure}{\textbf{Iteration:}}
\renewcommand{\algorithmiclastcon}{\textbf{Output:}} 

\newtheorem{assume}{Assumption}
\newcommand{\e}{\begin{equation}}
\newcommand{\ee}{\end{equation}}
\newcommand{\en}{\begin{equation*}}
\newcommand{\een}{\end{equation*}}
\newcommand{\eqn}{\begin{eqnarray}}
\newcommand{\eeqn}{\end{eqnarray}}
\newcommand{\bmat}{\begin{bmatrix}}
\newcommand{\emat}{\end{bmatrix}}
\newcommand{\BIT}{\begin{itemize}}
\newcommand{\EIT}{\end{itemize}}

\newcommand{\ve}{\bm e}

\newcommand{\vm}{\bm m}

\newcommand{\vq}{\bm q}

\newcommand{\vs}{\bm s}

\newcommand{\vu}{\bm u}

\newcommand{\vx}{\bm x}
\newcommand{\vy}{\bm y}
\newcommand{\vz}{\bm z}

\newcommand{\mB}{\bm B}

\newcommand{\mH}{\bm H}
\newcommand{\mI}{\bm I}

\newcounter{oursection}

\usepackage{color}
\usepackage{cleveref}
\usepackage{silence}
\usepackage{soul}
\begin{document}
%
\title{Solving RED with Weighted Proximal Methods}
%
%
%

\author{Tao Hong, Irad Yavneh, and Michael Zibulevsky
\thanks{T. Hong, I. Yavneh, and M. Zibulevsky are with the Department
of Computer Science, Technion-Israel Institute of Technology, Haifa, 32000 Israel. (Email: {\{hongtao,irad,mzib\}@cs.technion.ac.il}).}
}

%
%

\markboth{}
{Shell \MakeLowercase{\textit{et al.}}: Bare Demo of IEEEtran.cls for IEEE Journals}
%



\maketitle

\begin{abstract}
REgularization by Denoising (RED) is an attractive framework for solving inverse problems by incorporating state-of-the-art denoising algorithms as the priors. A drawback of this approach is the high computational complexity of denoisers, which dominate the computation time. In this paper, we apply a general framework called weighted proximal methods (WPMs) to solve RED efficiently. We first show that two recently introduced RED solvers (using the fixed point and accelerated proximal gradient methods) are particular cases of WPMs. Then we show by numerical experiments that slightly more sophisticated variants of WPM can lead to reduced run times for RED by requiring a significantly smaller number of calls to the denoiser. 
\end{abstract}

%
\IEEEpeerreviewmaketitle

\section{Introduction}
%
%
%
%

\IEEEPARstart{T}{HE}
goal of inverse problems is to recover an unknown signal $\vx\in\Re^N$ from an indirect measurement $\vy\in\Re^M$. The measurement is commonly modelled as  $\vy={\mathcal H}(\vx)+\ve,$ where $\mathcal H(\cdot)$ denotes an abstract operator and $\ve$ is often assumed to be white Gaussian noise with mean zero and variance $\sigma^2$. In this paper, we assume $\mathcal H(\cdot)$ to be a linear operator, $\mathcal H(\vx)=\mH\vx,$ with $\mH\in\Re^{M\times N}$, and focus on natural images. Lacking any prior knowledge about the signal $\vx$, we may reconstruct $\vx$ via the maximum likelihood (ML) minimization problem,
\e
\vx_{ML}^*=\arg\min_{\vx}\frac{1}{2}\|\mH\vx-\vy\|_2^2. \label{eq:MLEsti}
\ee
However, it is well-known that this approach is not generally useful. Even in the simple denoising problem, where $\mH$ is the identity matrix, ML results in $\vx^*_{ML}=\vy$, that is, we simply recover the noisy image. Furthermore, quite often $M<N$, resulting in infinitely many solutions, and even if this is not the case, $\mH$ may be highly ill-conditioned. For these reasons, the prevalent approach is to assume that the signal $\vx$ is sampled from some prior distribution, and to employ the {{maximum a posteriori probability (MAP) estimator}}, as formulated in \Cref{Sec:Prelim}. In our setting, MAP will result in adding to the right-hand side of (1) a term $\alpha R(\bm{x})$, where $\alpha$ is a parameter and $R$ is a regularization as discussed below. This approach has been applied with a large variety of priors, such as  $\ell_2$-based regularization \cite{king1984two}, wavelets \cite{chambolle1998nonlinear}, total variation \cite{rudin1992nonlinear}, kernel regularization \cite{zoran2011learning}, sparsity \cite{elad2006image}, and neural networks \cite{chen2017trainable}.

Naturally, the most widely studied problem in this framework is image denoising, e.g.,   \cite{buades2005non,elad2006image,dabov2007image,chen2017trainable,dong2013nonlocally}. Indeed, recent work suggests that the performance of leading image denoisers is close to a possible ceiling \cite{chatterjee2010denoising,milanfar2013tour,levin2011natural}. The availability of such powerful denoising algorithms has motivated researchers to seek ways to employ denoisers as priors for quite general inverse problems. The authors in \cite{protter2009generalizing,danielyan2012bm3d,metzler2015optimal} ``manually'' adopted priors used in existing denoisers for specific alternative inverse problems. Following this path, several authors proposed a general framework, called Plug-and-Play Priors ($P^3$) \cite{venkatakrishnan2013plug,sreehari2016plug}, for using the abundance of high-performance image denoisers as priors for other inverse problems. These authors formulate inverse problems as an optimization task and employ an Alternating Direction Method of Multipliers (ADMM) algorithm to tackle the corresponding minimization problem \cite{boyd2011distributed}. The image denoising algorithm is incorporated in each step of ADMM as an implicit prior.

Motivated by $P^3$, Romano et al. introduced REgularization by Denoising (RED) \cite{romano2017little}, which defines an optimization problem that includes the denoiser as an explicit prior. Given a differentiable denoiser $f(\vx)$, RED employs the following prior, 
\e
R(\vx)=\frac{1}{2}\vx^\mathcal T\left(\vx-f(\vx)\right), \label{eq:REDPrior}
\ee
where $\cdot^\mathcal T$ denotes transpose. Under two assumptions this leads to a convex minimization problem, and standard gradient based iterative methods are guaranteed to converge to a global minimum. Further details are provided in \Cref{Sec:Prelim}. 

Using state-of-the-art denoisers to construct priors is appealing, as it enables us to exploit the vast progress in denoising algorithms for addressing general inverse problems, and RED is a good framework to achieve this goal due to its flexibility. However, RED may be relatively expensive because at each iteration we must apply the denoising algorithm to evaluate the gradient, and the complexity of denoising algorithms is generally high. Indeed, the numerical experiments in \cite{romano2017little} reveal this concern. In that paper the authors propose three solvers for RED, namely, steepest descent (SD), the fixed-point (FP) method and the ADMM scheme. Amongst these, the FP method is the most efficient, but it still needs hundreds of iterations to complete the recovery process. 

Recently, \cite{hong2019acceleration} employed vector extrapolation to accelerate the FP method for RED, whereas \cite{reehorst2019regularization} applies an accelerated proximal gradient (APG) algorithm\footnote{APG is also known as FISTA \cite{beck2009fast} or Nesterov's acceleration \cite{nesterov2018lectures}.}. Both these approaches are faster than FP for RED, but they still require dozens of iterations. In this paper, we propose a general framework called weighted proximal methods (WPMs) \cite{beck2017first}. We show that FP and APG are in fact two particular variants of WPMs, and that by seeking a more effective weighting for WPMs we obtain a faster algorithm.
	
The rest of this paper is organized as follows. We review the RED framework and the FP and APG solvers in \Cref{Sec:Prelim}. The general WPM scheme is introduced in \Cref{Sec:WPMs}, and the choice of weighting is discussed. Numerical experiments on image deblurring and super-resolution tasks are presented in \Cref{Sec:NumExps} to demonstrate the efficiency of WPMs, followed by conclusions in \Cref{Sec:Conclusion}.

\section{REgularization by Denoising (RED)}\label{Sec:Prelim}
The MAP recovery process is formulated as follows:
$$
\begin{array}{rcl}
{\vx}^*_{MAP}& =& \arg\max_{\vx}P(\vx|\vy)~~~~~~~~~~~~~ (MAP)\\
&=&\arg\max_{\vx}P(\vy|\vx)P(\vx)~~~~~~~~ (Bayes~ rule)\\
&=&\arg\min_{\vx}-\log\{P(\vy|\vx)\}-\log{P(\vx)}.
\end{array}
$$
Assuming a robust Gibbs-like distribution of $\vx$, we have
$$
P(\vx)=constant \cdot e^{-\alpha R(\vx)},
$$
where $R$ denotes the so-called prior and $\alpha>0$ is a scaling parameter. Note that small $R(\vx)$ corresponds to highly probable signals. If $\ve$ is sampled from \emph{white Gaussian noise} with mean zero and variance $\sigma^2$, then we have 
$$
P(\vy|\vx)= constant \cdot e^{-\frac{1}{2\sigma^2}\|\mH\vx-\vy\|_2^2},
$$
This leads to the following minimization problem \cite{kirsch2011introduction},
\e
\vx^*_{MAP}=\arg\min_{\vx}~E(\vx) \triangleq \frac{1}{2\sigma^2} \|\mH\vx-\vy\|_2^2 + \alpha R(\vx). 
\label{eq:MAP:opt}
\ee

Substituting the RED prior \eqref{eq:REDPrior} into \eqref{eq:MAP:opt}, we obtain 
\e
\vx_{MAP}^*=\arg\min_{\vx} E(\vx)\triangleq \frac{1}{2\sigma^2}\|\mH\vx-\vy\|_2^2+\frac{\alpha}{2}\vx^\mathcal T\left(\vx-f(\vx)\right).\label{eq:opt:red:abs}
\ee

In \cite{romano2017little} two assumptions are made regarding the image denoising algorithm used in RED:
\begin{assume} \label{assume:local_homogeneity}
For any scalar $c$ arbitrarily close to $1$, $f(c\vx)=cf(\vx)$.
\end{assume}
\begin{assume}\label{assume:strong_passivity}
The spectral radius of the symmetric Jacobian $\nabla_{\vx}f(\vx)$ is upper bounded by one.
\end{assume}
Given Assumption \ref{assume:local_homogeneity}, we have

$$
\nabla_x f(\vx) \vx=\lim_{\epsilon \rightarrow 0}\frac{f(\vx+\epsilon \vx) -f(\vx)}{\epsilon}=f(\vx).
$$
Hence, the gradient of $R(\vx)$ is the residual of the denoiser,
\e
\nabla_{\vx} R(\vx)= \vx-f(\vx). \label{eq:opt:priors:gradient}
\ee
With \eqref{eq:opt:priors:gradient}, the gradient of $E(\vx)$ becomes
\e
\nabla_{\vx}E(\vx) = \frac{1}{\sigma^2}\mH^\mathcal T\left(\mH\vx-\vy\right)+\alpha \left(\vx-f(\vx)\right).\label{eq:opt:red:wholegradient}
\ee



Assumption \ref{assume:strong_passivity} implies convexity of $R(\vx)$, and therefore of $E(\vx)$ as well. Hence, any solution of $\nabla_{\vx}E(\vx) = 0$ yields a global minimum. This is a nonlinear problem, and we therefore resort to iterative solvers. One such solver is the FP method mentioned above, which lags the nonlinear term $f(\vx)$:
\e
\frac{1}{\sigma^2}\mH^\mathcal T\left(\mH\vx_{k+1}-\vy\right)+\alpha \left(\vx_{k+1}-f(\vx_k)\right)=\bm 0.\label{eq:FP:recursive:first-order}
\ee
We note that \eqref{eq:FP:recursive:first-order} can efficiently be solved for $\vx_{k+1}$ exactly in the Fourier domain if $\mH$ is block-circulant, or treated iteratively for a general $\mH$. The FP method can be accelerated using the APG approach as described in the following algorithm. Further discussion of APG can be found in \cite{reehorst2019regularization}. 

\begin{algorithm}[!htb]        
\caption{The APG Method \cite{reehorst2019regularization}}         
\label{alg:APG} 
\begin{algorithmic}[1]
\REQUIRE $\vx_0,$ $\vz_0=\vx_0$, and $t_0 = 1$.
\ENSURE 
\FOR {$k=0,1,\dots$}
\STATE Compute $\vx_{x+1}$ by solving \eqref{eq:FP:recursive:first-order}, with $\vz_k$ substituted for $\vx_k$ as the input from the last iteration 
\STATE $t_{k+1} = \frac{1+\sqrt{1+4t_k^2}}{2}$
\STATE $\vz_{k+1}\leftarrow \vx_{k+1}+\frac{t_k-1}{t_{k+1}}(\vx_{k+1}-\vx_k)$
\ENDFOR
\end{algorithmic}
\end{algorithm}



\section{Weighted Proximal Methods}\label{Sec:WPMs}

Consider the following composite problem and assume its solution set is nonempty,
\e
\min_{\vx} \mathcal F(\vx)\triangleq g(\vx)+h(\vx),\label{eq:compositemodel}
\ee
where $g$ and $h$ are convex and differentiable. Denote the proximal operator by
\e
\text{prox}_{h,\mB}(\hat{\vx})=\arg\min_{\vu} \left\{ h(\vu)+\frac{1}{2}\|\vu-\hat{\vx}\|_{\mB}^2 \right\},\label{eq:proxOper} 
\ee
where $\mB$ is a symmetric positive definite matrix called the weighting and $\|\cdot\|_{\mB}$ denotes the $\mB$-norm, $\|\vq\|_{\mB}=\sqrt{\vq^\mathcal T \mB \vq}$. With these, we describe the explicit form of WPMs for \eqref{eq:compositemodel} in \Cref{alg:WPMs} \cite[Chap. 10.7.5]{beck2017first}. Note that by setting $\mB=\beta\mI$ with $\beta>0$, we recover the proximal gradient (PG) method. Usually, PG is used for \eqref{eq:compositemodel} when $h$ is nonsmooth \cite{roy2018new}, whereas here we use it even though $h$ is differentiable. We do this for computational efficiency, knowing that applying the denoiser is the most expensive part of the solution process.
\begin{algorithm}[!htb]        
\caption{Weighed Proximal Methods (WPMs)}         
\label{alg:WPMs} 
\begin{algorithmic}[1]
\REQUIRE $\vx_0$.
\ENSURE 
\FOR {$k=0,1,\dots$}
\STATE pick the step-size $a_k$ and the weighting $\mB_k$
\STATE $\vx_{k+1}\leftarrow \text{prox}_{a_kh,\mB_k}\left(\vx_k-a_k\mB_k^{-1}\nabla_{\vx} g(\vx_k)\right)$
\ENDFOR
\end{algorithmic}
\end{algorithm}


To apply \Cref{alg:WPMs} to RED, we set $g(\vx)=\alpha R(\vx)=\frac{\alpha}{2}\vx^\mathcal T\left(\vx-f(\vx)\right)$ and $h(\vx)=\frac{1}{2\sigma^2}\|\mH\vx-\vy\|_2^2$. If $h(\vx)$ is convex, solving \eqref{eq:proxOper} is equivalent to satisfying the first-order optimality condition,
\e
\nabla_{\vu} h(\vu)+\mB(\vu-\hat{\vx})=\bm 0.\label{eq:ProxFirstOrder}
\ee
Substituting $\hat{\vx} \leftarrow \vx_k-a_k\mB_k^{-1}\nabla_{\vx} g(\vx_k)$, $h(\vu) \leftarrow a_k h(\vu)$ and $\vu \leftarrow \vx_{k+1}$, at the $k$th iteration into \eqref{eq:ProxFirstOrder} and rearranging, we obtain
\e
\left(\frac{a_k}{\sigma^2}\mH^\mathcal T\mH+\mB_k\right)\vx_{k+1} =\frac{a_k}{\sigma^2}\mH^\mathcal T\vy+\mB_k\vx_k-a_k\alpha\left(\vx_k-f(\vx_k)\right).\label{eq:proxOperLineq}
\ee
In this paper, we use the conjugate gradient (CG) method to approximately solve \eqref{eq:proxOperLineq} for $\vx_{k+1}$.

Next we discuss possible practical choices for the weighting $\mB_k$. Note first that if we set $\mB_k = \alpha \mI$, where $\mI$ is the identity matrix, and select the step-size $a_k=1$, \eqref{eq:proxOperLineq} is reduced to \eqref{eq:FP:recursive:first-order} and we recover the FP method. Moreover, by using the accelerated version of \Cref{alg:WPMs} (cf. \cite[Chap. 10.7.5]{beck2017first}) we get APG \cite{reehorst2019regularization} . We now propose a more elaborate approach of choosing some approximation to the Hessian of $\alpha R(\vx)$ as the weighting. (Because of the abstract denoiser in $R(\vx)$, the exact Hessian is not computable.) Specifically, we choose the \emph{symmetric-rank-one} (SR1) approximation to the Hessian \cite[Chap. 6.2]{jorge2006numerical}, as is used in quasi-Newton methods. The SR1 approximation is described in \Cref{alg:SR1}. This choice yields faster convergence in our experiments than either FP or APG, as shown below. We henceforth use WPM to denote \Cref{alg:WPMs} with the weighting chosen by \Cref{alg:SR1}.

\begin{algorithm}[!htb]        
\caption{SR1 updating}         
\label{alg:SR1} 
\begin{algorithmic}[1]
\REQUIRE $k=1$, $\gamma =1.25$, $\delta = 10^{-8}$, $\vx_k$, $\vx_{k-1}$, $\nabla g(\vx_k)$, $\nabla g(\vx_{k-1})$.
\IF {$k=1$}
\STATE $\mB_k\leftarrow\alpha\mI$
\ELSE
\STATE Set $\vs_k \leftarrow \vx_k-\vx_{k-1}$ and $\vm_k \leftarrow \nabla g(\vx_k)-\nabla g(\vx_{k-1})$
\STATE Calculate $\tau \leftarrow \gamma\frac{\|\vm_k\|_2^2}{\left<\vs_k,\vm_k\right>}$
\IF {$\tau<0$}
\STATE $\mB_k\leftarrow\alpha\mI$
\ELSE
\STATE $\mH_0 \leftarrow \tau\mI$
\IF {$|\left<\vm_k-\mH_0\vs_k,\vs_k\right>|\leq \delta \|\vs_k\|_2\|\vm_k-\mH_0\vs_k\|_2$}
\STATE $\vu_k\leftarrow \bm 0$
\ELSE
\STATE $\vu_k\leftarrow \frac{\vm_k-\mH_0\vs_k}{\sqrt{\left<\vm_k-\mH_0\vs_k,\vs_k\right>}}$
\ENDIF
\STATE $\mB_k \leftarrow \mH_0+\vu_k\vu_k^\mathcal T$\label{alg:SR1:HessUpdat}
\ENDIF
\ENDIF
\STATE {\bf Return:} $\mB_k$
\end{algorithmic}
\end{algorithm}

Unlike the traditional SR1, we formulate each $\mB_k$ from the initial $\mH_0$ rather the previous iterate $\mB_{k-1}$ \cite{jorge2006numerical}. Moreover, we scale $\mH_0$ by $\gamma>1$ as suggested in \cite{becker2012quasi}, which we found useful in practice. In the practical implementation of \Cref{alg:SR1}, we efficiently represent $\mB_k$ as a matrix-vector multiplication operator rather than as an explicit matrix.

In general, the step-size $a_k$ in \Cref{alg:WPMs} needs to be chosen by some line search process to guarantee monotonically decreasing objective values at each iteration. However, because evaluating the objective value in RED requires calling the denoiser, standard line search methods may dramatically increase the complexity of the algorithm. To maintain a low computational cost, we fix $a_k=1$ and reduce the step-size by half only if the objective value exhibits a relative growth above some threshold, i.e., $E(\vx_{k+1})-E(\vx_k)>\varepsilon E(\vx_{k+1})$, where we use $\varepsilon=10^{-2}$ in all our experiments. In practice, we found that we never needed to reduce the step-size. 

In this paper we only investigate the SR1 approximation to the Hessian of $\alpha R(\vx)$. We acknowledge that a more accurate Hessian estimate may prove to be even more cost-effective for RED, but leave such investigation to future work. Because we use an approximate Hessian for the weighting, our algorithm is equivalent to a quasi-newton proximal method. It follows that if both $g(\vx)$ and $h(\vx)$ are strongly convex and their gradients are Lipschitz continuous, WPM with SR1 estimation, an appropriate step-size $a_k$, and exact solution of \eqref{eq:proxOper}, converges linearly; see details in \cite{becker2019quasi}. Because we depart from these strict requirements for efficiency, we cannot claim provable convergence in our implementation. However, in all our experiments WPM converged. Finally, we note that \cite{reehorst2019regularization} challenges the validity in practice of the underlying assumptions of RED for most denoisers, concluding that \eqref{eq:opt:red:wholegradient} is not truly the gradient of \eqref{eq:opt:red:abs}. Nevertheless, setting \eqref{eq:opt:red:wholegradient} to zero, as is the objective of all the algorithms we discuss here, remains a most attractive method for signal recovery.

\section{Numerical Experiments}\label{Sec:NumExps}
In this section we investigate the performance of solvers for RED. Following  \cite{romano2017little}, we perform our tests on image deblurring and super-resolution tasks and use the trainable nonlinear reaction diffusion (TNRD) \cite{chen2017trainable} method as the abstract denoiser. We remark that one can adopt deep denoising techniques instead of TNRD, since the differentiability requirement of the denoiser is not mandatory in practice \cite{reehorst2019regularization}.  This may possibly lead to improved results in practice, but we do not investigate such options here. Also, since the authors in \cite{romano2017little} already show the superiority of RED for image deblurring and super-resolution tasks compared with other popular algorithms, we largely omit such comparisons in this paper and concentrate on computational efficiency. Moreover, the experiments conducted in \cite{hong2019acceleration} demonstrated that the FP method converges faster than LBFGS and Nesterov's acceleration for RED. Therefore, we only compare WPM  to FP \cite{romano2017little}, FP-MPE \cite{hong2019acceleration}, and APG \cite{reehorst2019regularization}. All of the experiments are carried out on a laptop with Intel i$7-6500$U CPU @2.50GHz and 8GB RAM.

\begin{figure}[!htb]
    \centering
    \subfigure[Deblurring with uniform Blur.]{\includegraphics[scale=0.3]{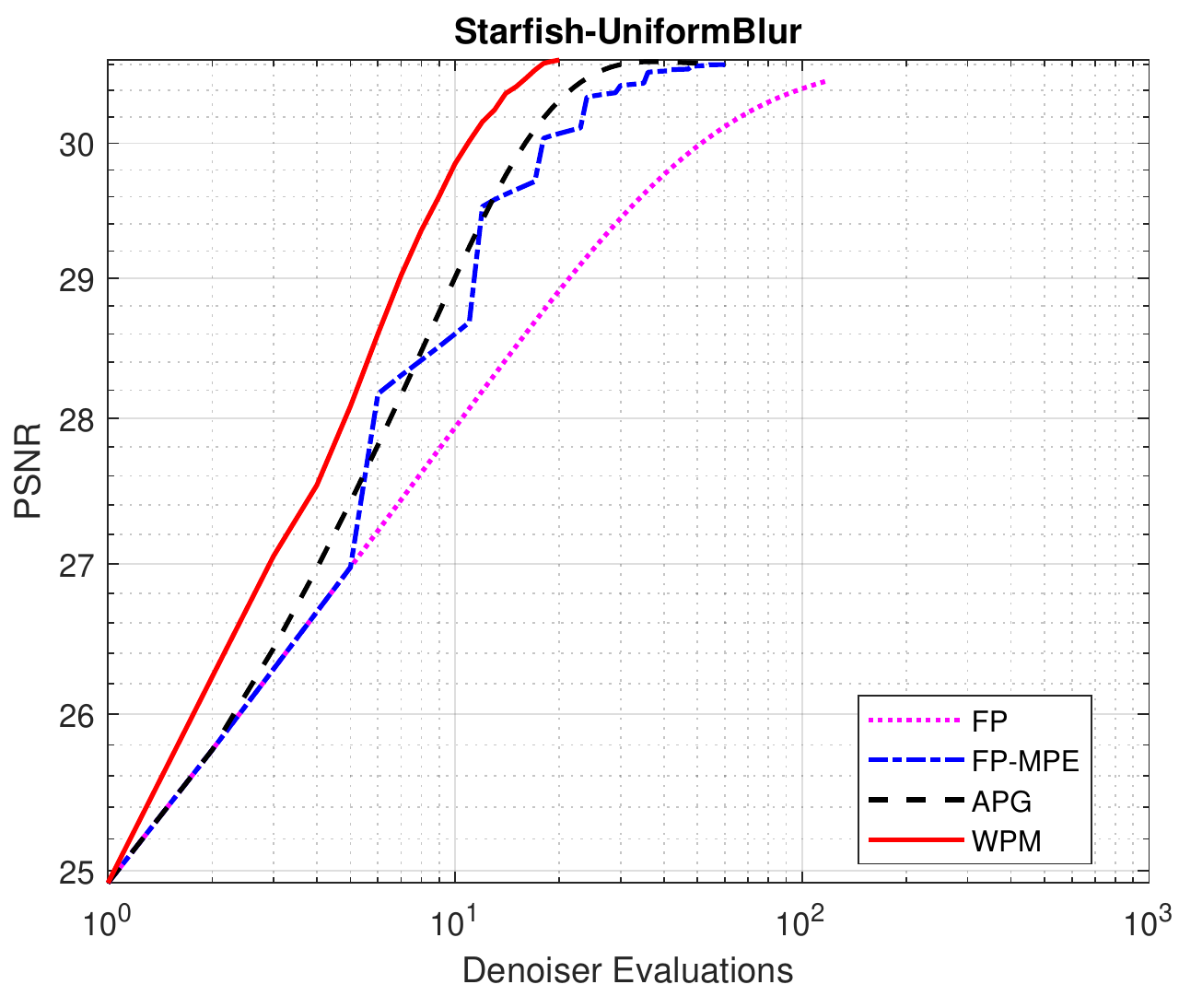}
    \includegraphics[scale=0.3]{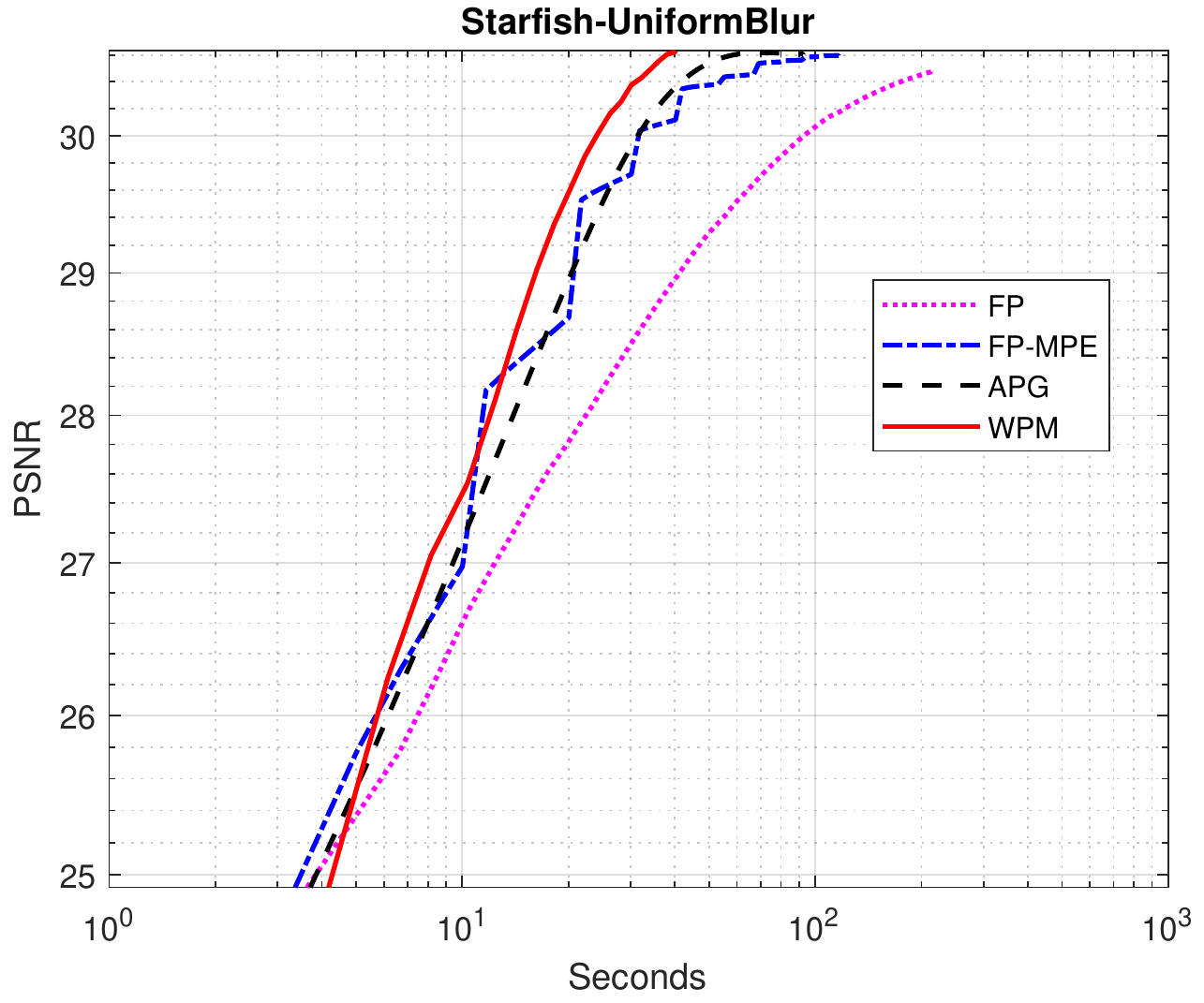}}
    
    \subfigure[Deblurring with Gaussian Blur.]{\includegraphics[scale=0.29]{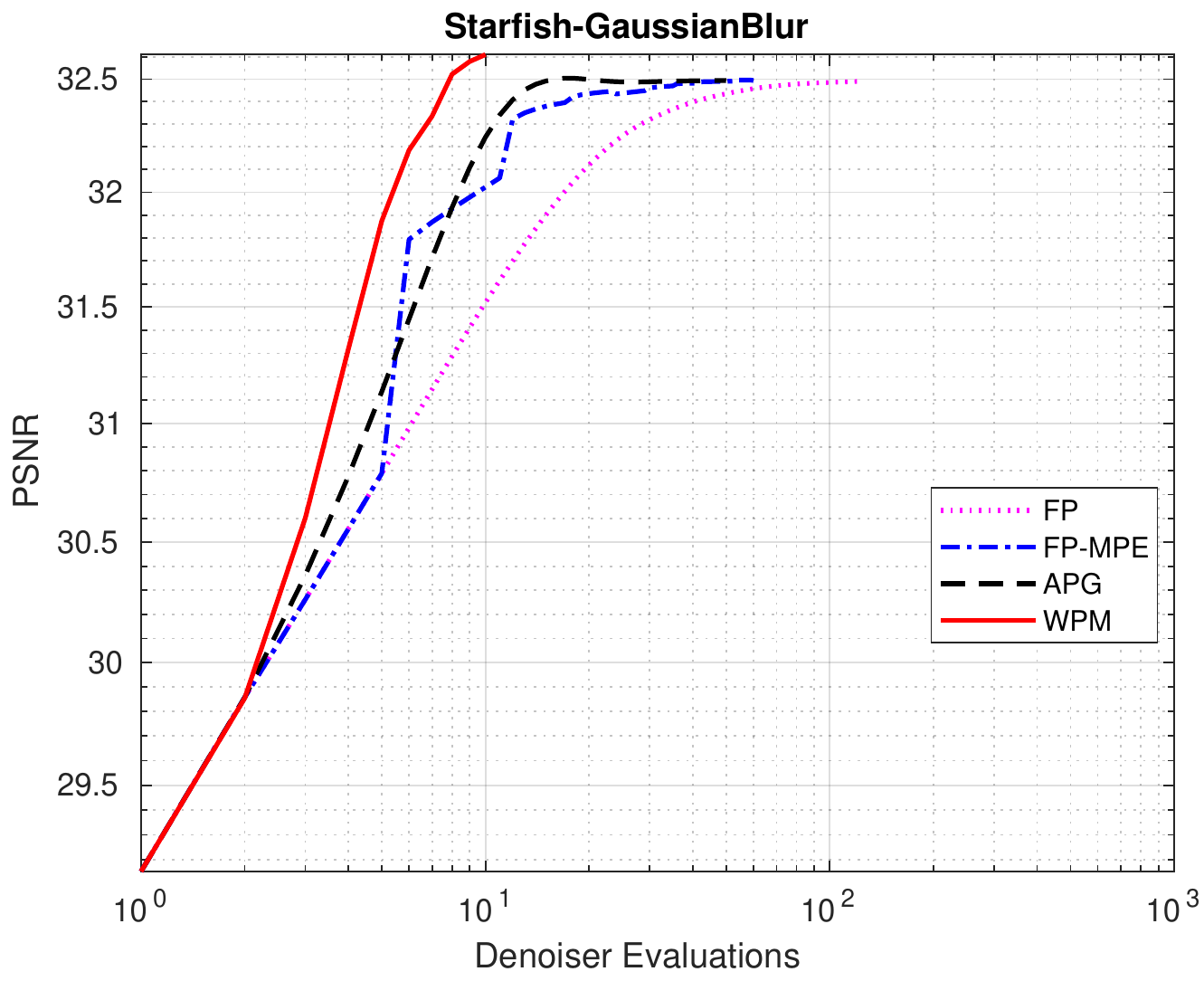}
    \includegraphics[scale=0.29]{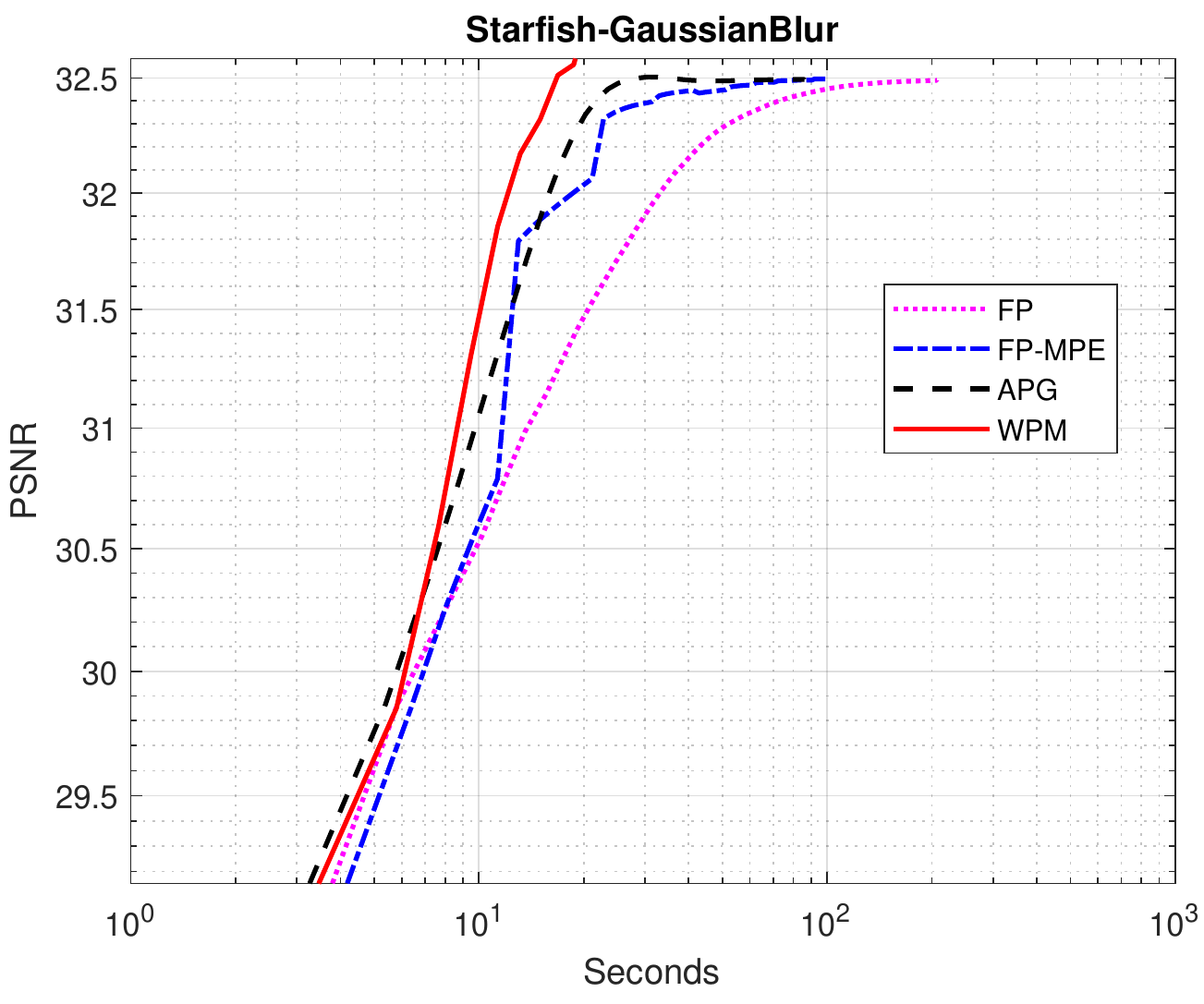}}
    \caption{PSNR versus denoiser evaluations (left column) and CPU time (right column) for deblurring the ``Starfish'' image.}
    \label{fig:deblurCompare}
\end{figure}

For image deblurring, the image is degraded by convolving with a point spread function (PSF), $9\times 9$ uniform blur or a Gaussian blur with a standard derivation $1.6$, and then adding Gaussian noise with mean zero and $\sigma = \sqrt{2}$. The recovered peak-signal-to-noise ratio (PSNR) versus the number of denoiser evaluations (left column) and running time (right column) when using RED for the ``Starfish'' image are shown in \Cref{fig:deblurCompare}. We find that the performances of FP-MPE and APG are similar, whereas WPM is more efficient than both, requiring less denoiser evaluations and running time to achieve a comparable PSNR. 
These results also indicate that indeed the denoiser dominates the complexity of solving RED. 

\begin{figure}[!htb]
    \centering
    {
    \includegraphics[scale=0.3]{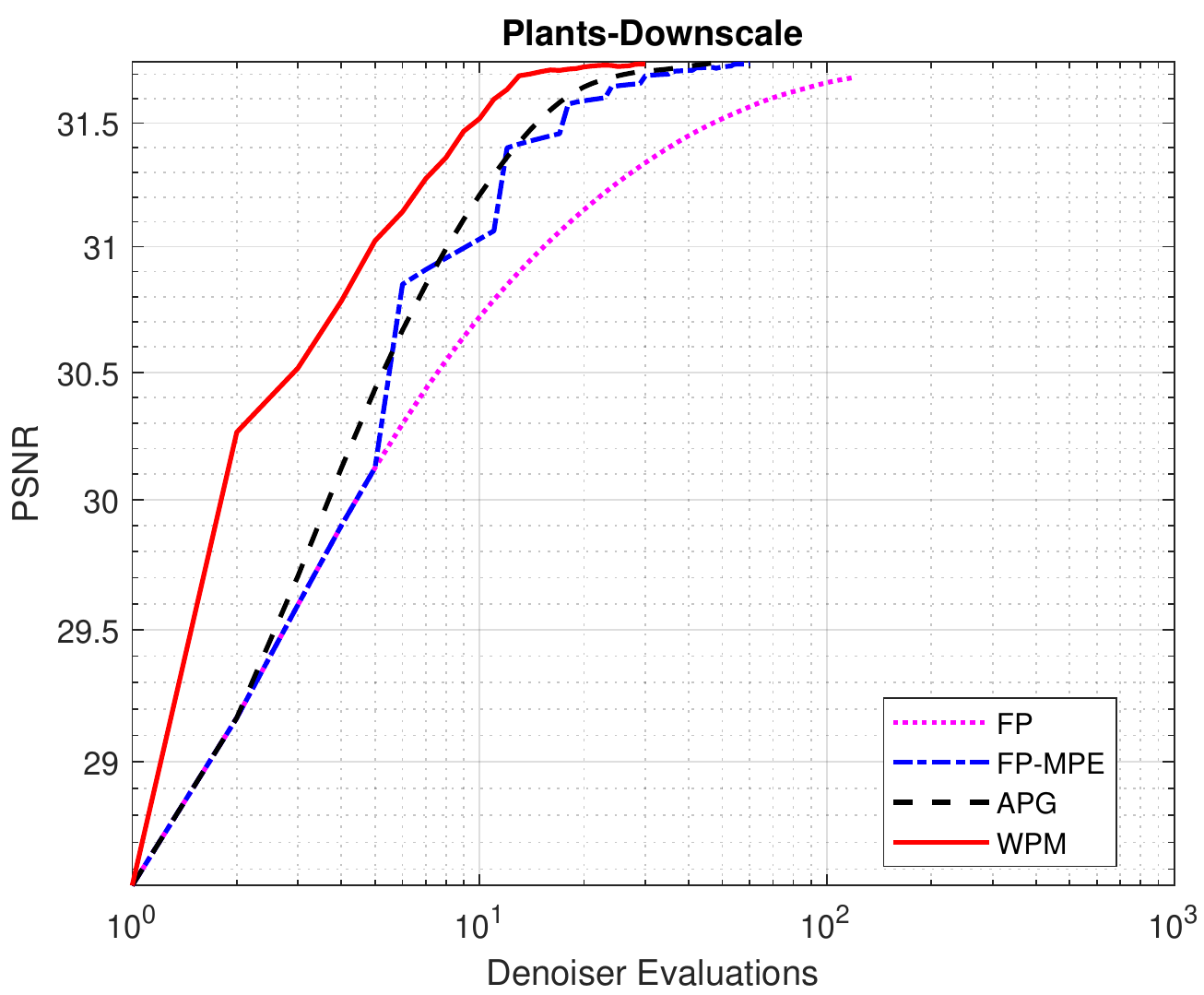}
    \includegraphics[scale=0.3]{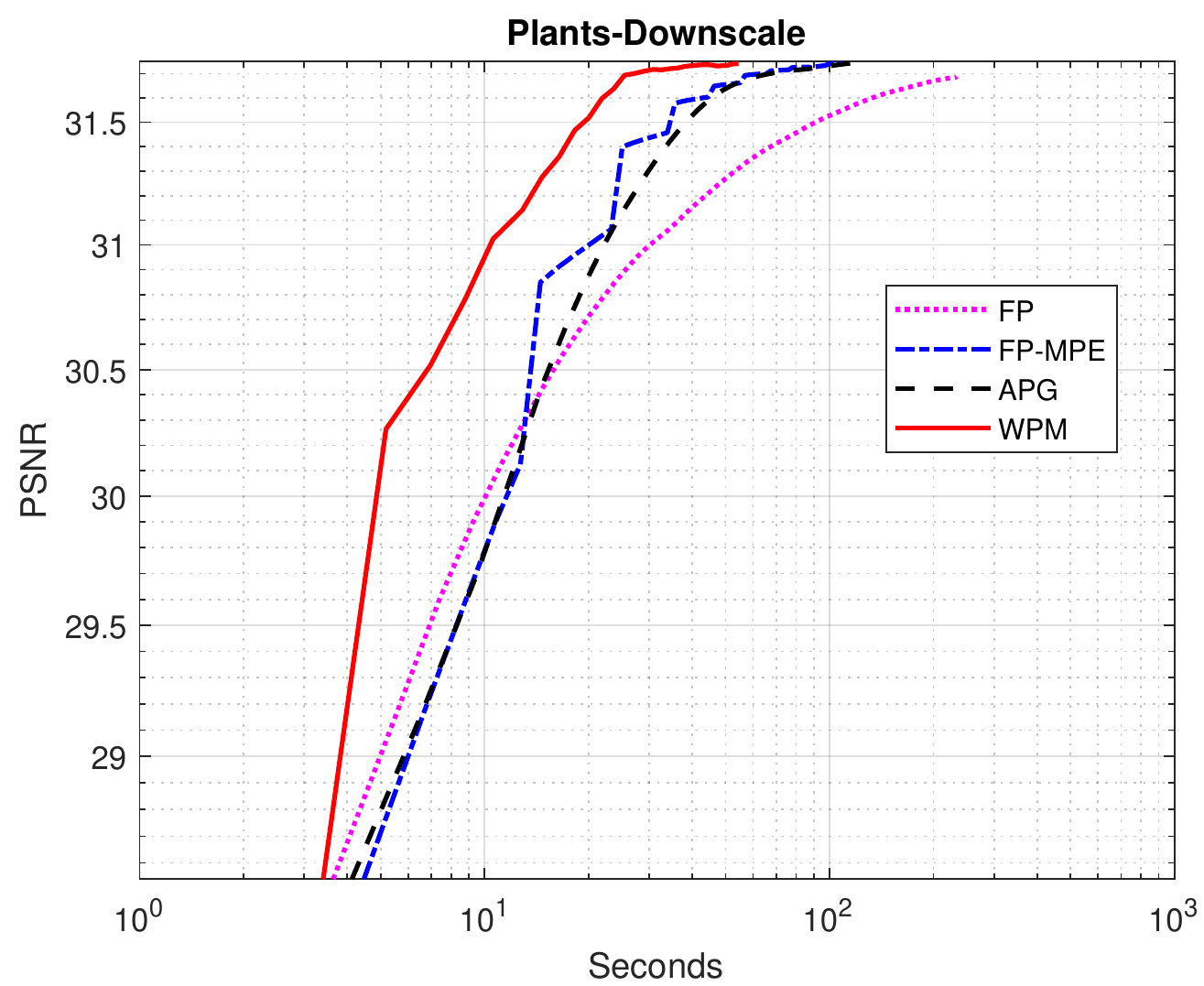}
    }
    \caption{PSNR versus denoiser evaluations (left) and CPU time (right) for super-resolution of the ``Plants'' image.}
    \label{fig:SRCompare}
\end{figure}

Next, we test the algorithms on image super-resolution. A low resolution image is generated by blurring a high-resolution image with a $7\times 7$ Gaussian kernel with standard derivation $1.6$, and then downscaling by a factor of 3. To the resulting image we add Gaussian noise with mean zero and $\sigma=5$, resulting in our deteriorated image. The PSNR of the recovered fine-resolution image versus the number of denoiser evaluations (left) and running time (right) for the ``Plants'' image are presented in \Cref{fig:SRCompare}. Again, we observe that WPM requires less denoiser evaluations and running time to achieve a comparable PSNR. 

\begin{figure}
    \centering
    {\small Deblurring -- Uniform}
    
    \subfigure[Original]{\includegraphics[scale=0.152]{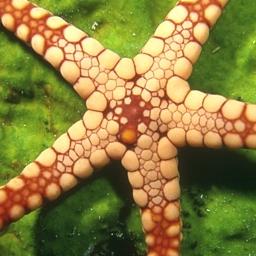}}
    \subfigure[Blurred]{\includegraphics[scale=0.15]{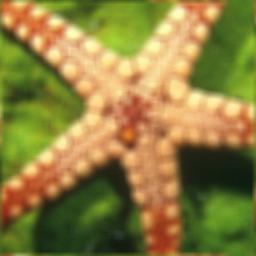}}
    \subfigure[27.94]{\includegraphics[scale=0.15]{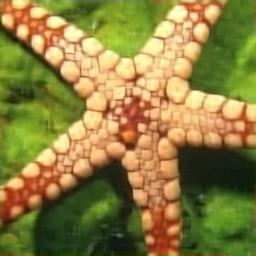}}
    \subfigure[28.60]{\includegraphics[scale=0.15]{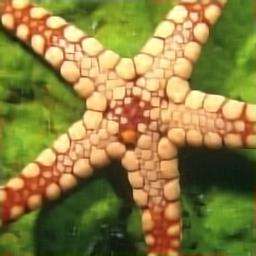}}
    \subfigure[29.01]{\includegraphics[scale=0.15]{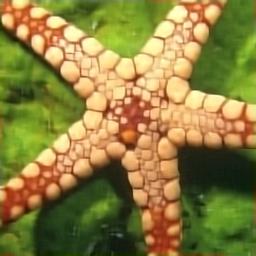}}
    \subfigure[29.85]{\includegraphics[scale=0.15]{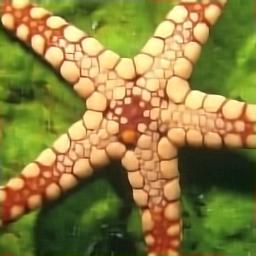}}
    
    {\small Deblurring -- Gaussian}
    
    \subfigure[Original]{\includegraphics[scale=0.152]{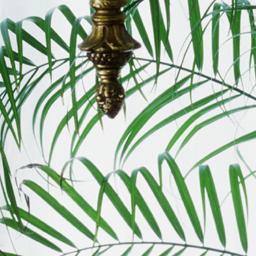}}
    \subfigure[Blurred]{\includegraphics[scale=0.15]{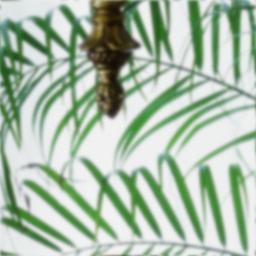}}
    \subfigure[30.13]{\includegraphics[scale=0.15]{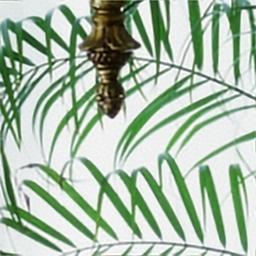}}
    \subfigure[30.91]{\includegraphics[scale=0.15]{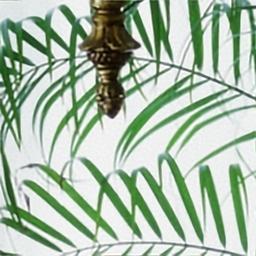}}
    \subfigure[31.23]{\includegraphics[scale=0.15]{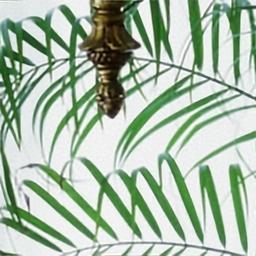}}
    \subfigure[31.63]{\includegraphics[scale=0.15]{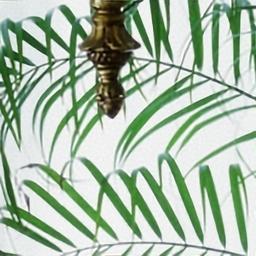}}

    {\small Super-Resolution}

    \subfigure[Original]{\includegraphics[scale=0.161]{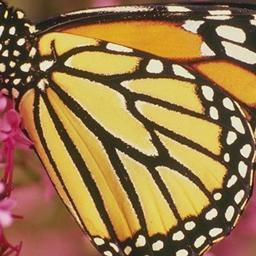}}
    \subfigure[LR]{\hspace{0.5cm}\includegraphics[scale=0.15]{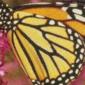}\hspace{0.5cm}}
    \subfigure[24.56]{\includegraphics[scale=0.15]{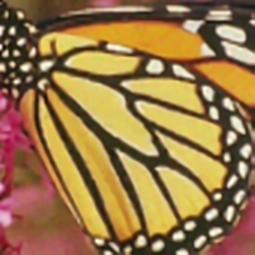}}
    \subfigure[25.13]{\includegraphics[scale=0.15]{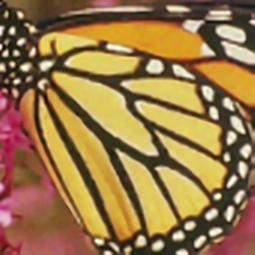}}
    \subfigure[25.38]{\includegraphics[scale=0.15]{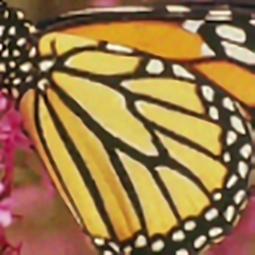}}
    \subfigure[26.20]{\includegraphics[scale=0.15]{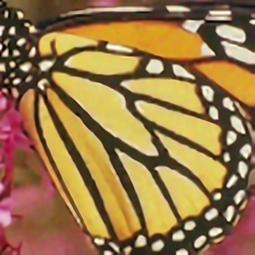}}  
    
    \caption{PSNR (dB) of the image recovered by, from left to right, FP, FP-MPE, APG, and WPM, after $10$ denoiser evaluations. LR stands for low-resolution.}
    \label{fig:VisualEffect}
\end{figure}

Examining the performance of the algorithms further, we run them on eight additional images tested in \cite{romano2017little}. For each image, we run the FP method with $200$ denoiser evaluations and take the final PSNR as a benchmark. Then we examine how many denoiser evaluations are needed for APG, FP-MPE and WPM, to achieve a similar PSNR. The results are listed in \Cref{tab:deblur:psnr}. Evidently, with the exception of ``Boats'' and ``House'' in the deblurring task, we observe that WPM requires the smallest number of denoiser evaluations to achieve a comparable PSNR, demonstrating its efficiency for solving RED. Additionally, we present the recovered results of the ``Starfish'' and ``Leaves'' images from deblurring with uniform and Gaussian blurs, respectively, and the ``Butterfly'' image from super-resolution in \Cref{fig:VisualEffect} to visualize the effectiveness of RED solved by WPM.
%

\begin{table}[!htb]
		\centering
		\caption{Denoiser evaluations required to attain a similar PSNR. The first and second rows per each image refer to image deblurring and the third row refers to super-resolution. The minimal number of denoiser evaluations is marked in bold.}
		
		\begin{tabular}{c||c|c|c}
		\hline
		     & FP-MPE & APG & WPM \\
		     \hline 
		     \hline
		     \multirow{3}{*}{\textsf{Butterfly}}& $54$ & $34$ & $\bm{25}$\\
		                                        & $54$ & $26$ & $\bm{17}$\\
		                                        & $80$ & $51$ & $\bm{26}$\\
		                                        \hline
		     \multirow{3}{*}{\textsf{Boats}}   &  $24$ & $\bm{20}$ & $21$\\
		                                       &  $60$ & $34$ & $\bm{22}$\\
		                                       & $36$ & $20$ & $\bm{12}$\\
		                                       \hline
		     \multirow{3}{*}{\textsf{House}}   & $24$ &  $\bm{18}$ & $19$\\
		                                       & $62$ &  $26$ & $\bm{25}$\\
		                                       & $18$& $15$ & $\bm{10}$\\
		                                       \hline
		     \multirow{3}{*}{\textsf{Parrot}} & $39$ & $30$ & $\bm{20}$\\
		                                      & $52$ & $40$ & $\bm{36}$\\
		                                      & $49$ & $31$ & $\bm{28}$\\
		                                      \hline
		     \multirow{3}{*}{\textsf{Lena}}  & $48$ & $34$ & $\bm{29}$\\
		                                     & $47$ & $16$ & $\bm{15}$\\
		                                     & $37$ & $26$ & $\bm{18}$\\
		                                     \hline
		     \multirow{3}{*}{\textsf{Barbara}} & $14$ & $12$ & $\bm{11}$\\
		                                       & $48$ & $23$ & $\bm{16}$\\
		                                       & $17$ & $15$ & $\bm{11}$ \\
		                                      \hline
		     \multirow{3}{*}{\textsf{Peppers}} & $42$ & $29$ & $\bm{22}$\\
		                                       & $41$ & $40$ & $\bm{34}$\\
		                                       & $38$ & $30$ & $\bm{28}$\\
		                                       \hline
		     \multirow{3}{*}{\textsf{Leaves}} & $50$ & $41$ & $\bm{34}$\\
		                                      & $36$ & $18$ & $\bm{14}$\\
		                                      & $60$ & $41$ & $\bm{12}$\\
 		    \hline
		\end{tabular}

		\label{tab:deblur:psnr}
\end{table}

\section{Conclusion} \label{Sec:Conclusion}
In this paper, we propose a general framework for RED called weighted proximal methods (WPMs). By setting $\mB_k=\alpha \mI$ and $a_k=1$, we retrieve the FP and APG methods. However, by choosing the weighting to be an approximation to the Hessian of $\alpha R(\vx)$, we obtain a more efficient algorithm. The experiments on image deblurring and super-resolution tasks demonstrate that WPM with a simple and inexpensive approximation to the Hessian can substantially reduce the overall number of denoiser evaluations in the recovery process, usually resulting in significant speedup. In future work we aim to design better Hessian approximations in order to accelerate the computation further.

\ifCLASSOPTIONcaptionsoff
  \newpage
\fi

\newpage
\bibliographystyle{IEEEtran}
\bibliography{Refs}

\begin{thebibliography}{10}
\providecommand{\url}[1]{#1}
\csname url@samestyle\endcsname
\providecommand{\newblock}{\relax}
\providecommand{\bibinfo}[2]{#2}
\providecommand{\BIBentrySTDinterwordspacing}{\spaceskip=0pt\relax}
\providecommand{\BIBentryALTinterwordstretchfactor}{4}
\providecommand{\BIBentryALTinterwordspacing}{\spaceskip=\fontdimen2\font plus
\BIBentryALTinterwordstretchfactor\fontdimen3\font minus
  \fontdimen4\font\relax}
\providecommand{\BIBforeignlanguage}[2]{{%
\expandafter\ifx\csname l@#1\endcsname\relax
\typeout{** WARNING: IEEEtran.bst: No hyphenation pattern has been}%
\typeout{** loaded for the language `#1'. Using the pattern for}%
\typeout{** the default language instead.}%
\else
\language=\csname l@#1\endcsname
\fi
#2}}
\providecommand{\BIBdecl}{\relax}
\BIBdecl

\bibitem{king1984two}
M.~A. King, R.~B. Schwinger, P.~W. Doherty, and B.~C. Penney, ``Two-dimensional
  filtering of spect images using the metz and wiener filters,'' \emph{Journal
  of Nuclear Medicine}, vol.~25, no.~11, pp. 1234--1240, 1984.

\bibitem{chambolle1998nonlinear}
A.~Chambolle, R.~A. De~Vore, N.-Y. Lee, and B.~J. Lucier, ``Nonlinear wavelet
  image processing: variational problems, compression, and noise removal
  through wavelet shrinkage,'' \emph{IEEE Transactions on Image Processing},
  vol.~7, no.~3, pp. 319--335, 1998.

\bibitem{rudin1992nonlinear}
L.~I. Rudin, S.~Osher, and E.~Fatemi, ``Nonlinear total variation based noise
  removal algorithms,'' \emph{Physica D: nonlinear phenomena}, vol.~60, no.
  1-4, pp. 259--268, 1992.

\bibitem{zoran2011learning}
D.~Zoran and Y.~Weiss, ``From learning models of natural image patches to whole
  image restoration,'' in \emph{Computer Vision (ICCV), 2011 IEEE International
  Conference on}.\hskip 1em plus 0.5em minus 0.4em\relax IEEE, 2011, pp.
  479--486.

\bibitem{elad2006image}
M.~Elad and M.~Aharon, ``Image denoising via sparse and redundant
  representations over learned dictionaries,'' \emph{IEEE Transactions on Image
  processing}, vol.~15, no.~12, pp. 3736--3745, 2006.

\bibitem{chen2017trainable}
Y.~Chen and T.~Pock, ``Trainable nonlinear reaction diffusion: A flexible
  framework for fast and effective image restoration,'' \emph{IEEE Transactions
  on Pattern Analysis and Machine Intelligence}, vol.~39, no.~6, pp.
  1256--1272, 2017.

\bibitem{buades2005non}
A.~Buades, B.~Coll, and J.-M. Morel, ``A non-local algorithm for image
  denoising,'' in \emph{Computer Vision and Pattern Recognition, CVPR, IEEE
  Computer Society Conference on}, vol.~2, 2005, pp. 60--65.

\bibitem{dabov2007image}
K.~Dabov, A.~Foi, V.~Katkovnik, and K.~Egiazarian, ``Image denoising by sparse
  3-d transform-domain collaborative filtering,'' \emph{IEEE Transactions on
  Image Processing}, vol.~16, no.~8, pp. 2080--2095, 2007.

\bibitem{dong2013nonlocally}
W.~Dong, L.~Zhang, G.~Shi, and X.~Li, ``Nonlocally centralized sparse
  representation for image restoration,'' \emph{IEEE Transactions on Image
  Processing}, vol.~22, no.~4, pp. 1620--1630, 2013.

\bibitem{chatterjee2010denoising}
P.~Chatterjee and P.~Milanfar, ``Is denoising dead?'' \emph{IEEE Transactions
  on Image Processing}, vol.~19, no.~4, pp. 895--911, 2010.

\bibitem{milanfar2013tour}
P.~Milanfar, ``A tour of modern image filtering: New insights and methods, both
  practical and theoretical,'' \emph{IEEE Signal Processing Magazine}, vol.~30,
  no.~1, pp. 106--128, 2013.

\bibitem{levin2011natural}
A.~Levin and B.~Nadler, ``Natural image denoising: Optimality and inherent
  bounds,'' in \emph{Computer Vision and Pattern Recognition (CVPR), 2011 IEEE
  Conference on}.\hskip 1em plus 0.5em minus 0.4em\relax IEEE, 2011, pp.
  2833--2840.

\bibitem{protter2009generalizing}
M.~Protter, M.~Elad, H.~Takeda, and P.~Milanfar, ``Generalizing the
  nonlocal-means to super-resolution reconstruction,'' \emph{IEEE Transactions
  on Image Processing}, vol.~18, no.~1, pp. 36--51, 2009.

\bibitem{danielyan2012bm3d}
A.~Danielyan, V.~Katkovnik, and K.~Egiazarian, ``Bm3d frames and variational
  image deblurring,'' \emph{IEEE Transactions on Image Processing}, vol.~21,
  no.~4, pp. 1715--1728, 2012.

\bibitem{metzler2015optimal}
C.~A. Metzler, A.~Maleki, and R.~G. Baraniuk, ``Optimal recovery from
  compressive measurements via denoising-based approximate message passing,''
  in \emph{Sampling Theory and Applications (SampTA), 2015 International
  Conference on}.\hskip 1em plus 0.5em minus 0.4em\relax IEEE, 2015, pp.
  508--512.

\bibitem{venkatakrishnan2013plug}
S.~V. Venkatakrishnan, C.~A. Bouman, and B.~Wohlberg, ``Plug-and-play priors
  for model based reconstruction,'' in \emph{Global Conference on Signal and
  Information Processing (GlobalSIP), 2013 IEEE}.\hskip 1em plus 0.5em minus
  0.4em\relax IEEE, 2013, pp. 945--948.

\bibitem{sreehari2016plug}
S.~Sreehari, S.~V. Venkatakrishnan, B.~Wohlberg, G.~T. Buzzard, L.~F. Drummy,
  J.~P. Simmons, and C.~A. Bouman, ``Plug-and-play priors for bright field
  electron tomography and sparse interpolation,'' \emph{IEEE Transactions on
  Computational Imaging}, vol.~2, no.~4, pp. 408--423, 2016.

\bibitem{boyd2011distributed}
S.~Boyd, N.~Parikh, E.~Chu, B.~Peleato, J.~Eckstein \emph{et~al.},
  ``Distributed optimization and statistical learning via the alternating
  direction method of multipliers,'' \emph{Foundations and
  Trends{\textregistered} in Machine Learning}, vol.~3, no.~1, pp. 1--122,
  2011.

\bibitem{romano2017little}
Y.~Romano, M.~Elad, and P.~Milanfar, ``The little engine that could:
  Regularization by denoising (red),'' \emph{SIAM Journal on Imaging Sciences},
  vol.~10, no.~4, pp. 1804--1844, 2017.

\bibitem{hong2019acceleration}
T.~Hong, Y.~Romano, and M.~Elad, ``Acceleration of red via vector
  extrapolation,'' \emph{Journal of Visual Communication and Image
  Representation}, p. 102575, 2019.

\bibitem{reehorst2019regularization}
E.~T. Reehorst and P.~Schniter, ``Regularization by denoising: Clarifications
  and new interpretations,'' \emph{IEEE Transactions on Computational Imaging},
  vol.~5, no.~1, pp. 52--67, 2019.

\bibitem{beck2009fast}
A.~Beck and M.~Teboulle, ``A fast iterative shrinkage-thresholding algorithm
  for linear inverse problems,'' \emph{SIAM journal on imaging sciences},
  vol.~2, no.~1, pp. 183--202, 2009.

\bibitem{nesterov2018lectures}
Y.~Nesterov, \emph{Lectures on Convex Optimization}.\hskip 1em plus 0.5em minus
  0.4em\relax Springer, 2018.

\bibitem{beck2017first}
A.~Beck, \emph{First-Order Methods in Optimization}.\hskip 1em plus 0.5em minus
  0.4em\relax SIAM, 2017, vol.~25.

\bibitem{kirsch2011introduction}
A.~Kirsch, \emph{An introduction to the mathematical theory of inverse
  problems}.\hskip 1em plus 0.5em minus 0.4em\relax Springer Science \&
  Business Media, 2011, vol. 120.

\bibitem{roy2018new}
S.~Roy and A.~Borz{\`\i}, ``A new optimization approach to sparse
  reconstruction of log-conductivity in acousto-electric tomography,''
  \emph{SIAM Journal on Imaging Sciences}, vol.~11, no.~2, pp. 1759--1784,
  2018.

\bibitem{jorge2006numerical}
J.~Nocedal and S.~J. Wright, \emph{Numerical Optimization.}\hskip 1em plus
  0.5em minus 0.4em\relax Springer, 2006.

\bibitem{becker2012quasi}
S.~Becker and J.~Fadili, ``A quasi-newton proximal splitting method,'' in
  \emph{Advances in Neural Information Processing Systems}, 2012, pp.
  2618--2626.

\bibitem{becker2019quasi}
S.~Becker, J.~Fadili, and P.~Ochs, ``On quasi-newton forward-backward
  splitting: Proximal calculus and convergence,'' \emph{SIAM Journal on
  Optimization}, vol.~29, no.~4, pp. 2445--2481, 2019.

\end{thebibliography}

\end{document}